\documentclass[12pt]{article}

\usepackage[utf8]{inputenc}
\usepackage[T2A]{fontenc}
\usepackage[english]{babel}

\pdfoutput=1

\newcommand{\MSbar}{\overline{\rm MS}}
\newcommand{\DRbar}{\overline{\rm DR}}

\begin{document}
\title{Exact $\beta$-function in Abelian and non-Abelian $\mathcal{N}=1$ supersymmetric gauge models and its analogy
with QCD $\beta$-function in C-scheme.}

\author{
I.~O.~Goriachuk\footnote{io.gorjachuk@physics.msu.ru}\\
{\small{\em Moscow State University,}}\\
{\small{\em Faculty of Physics, Department of Theoretical Physics,}}\\
{\small{\em 119991, Moscow, Russia}},\\
\\
A.~L.~Kataev\footnote{kataev@ms2.inr.ac.ru},\\
{\small{\em Institute for Nuclear Research of the Russian Academy of Science,}}\\
{\small{\em 117312, Moscow, Russia}};\\
{\small{\em Moscow Institute of Physics and Technology,}}\\
{\small{\em 141700, Dolgoprudny, Moscow Region, Russia}}.\\
}

\maketitle
\vspace*{-14.0cm}
\begin{flushright}
INR-TH-2020-017
\end{flushright}
\vspace*{13.0cm}

\begin{abstract}
For $\mathcal{N}=1$ supersymmetric Yang-Mills theory without matter it is demonstrated that
there is a class of renormalization schemes, in which the exact Novikov, Shifman, Vainshtein, and Zakharov (NSVZ) formula
for the renormalization group $\beta$-function, defined in terms of the renormalized coupling constant, is valid.
These schemes are related with each other by finite renormalizations forming a one-parameter commutative subgroup
of general renormalization group transformations.
The analogy between the exact $\beta$-function in $\mathcal{N}=1$ supersymmetric Yang-Mills theory without matter
and the $\beta$-function of quantum chromodynamics in the C-scheme is discussed.
\end{abstract}

\textbf{1.}
According to the renormalization group method \cite{Bogolyubov:1980nc}, the evolution of the renormalized
coupling constant with the change of scale $\mu$ is described by the $\beta$-function:
\begin{equation}
\beta\left(a_s\right) =
\left.\frac{da_s\left({a_s}_0, {\mu}^2/{\Lambda}^2\right)}{d\ln{{\mu}^2}}\right\arrowvert_{{a_s}_0} =
- \left(\beta_{0} {a_s}^2 + \beta_{1} {a_s}^3 + \beta_{2} {a_s}^4 + O\left({a_s}^5\right)\right), \label{eq:beta_nA}
\end{equation}
where $a_s \equiv \alpha_s/\pi$, ${a_s}_0$ is the bare coupling constant and $\Lambda$ is the dimensionful parameter,
introduced in the theory by regularization.
The definition (\ref{eq:beta_nA}) is written in terms of the renormalized coupling constant.
The $\beta$-function coefficients $\beta_{i}$ (for $i \geq 2$) depend on the renormalization procedure.
They may be changed by the finite renormalization of the coupling constant.

The characteristic feature of $\mathcal{N}=1$ supersymmetric (SUSY) gauge theories is the existence of
the exact $\beta$-functions \cite{Novikov:1983uc}.
For example, in $\mathcal{N}=1$ supersymmetric quantum electrodynamics (SQED) with $N_f$ flavours of matter superfields
the corresponding exact $\beta$-function, expressed in terms of the renormalized coupling constant, reads
\cite{Vainshtein:1986ja}:
\begin{equation}
\beta\left(a\right) = a^2 N_f \left(\frac{1}{2} - \gamma\left(a\right)\right). \label{eq:NSVZ_SQED}
\end{equation}
This formula relates the $\beta$-function and the anomalous dimension of matter
\begin{equation}
\gamma\left(a\right) = \left.\frac{d\ln{Z}\left(a\left(a_0, {\mu}^2/{\Lambda}^2\right),
{\mu}^2/{\Lambda}^2\right)}{d\ln{{\mu}^2}}\right\arrowvert_{a_0},
\end{equation}
where $Z\left(a, {\mu}^2/{\Lambda}^2\right)$ is the renormalization  constant of matter superfields.
It was shown in \cite{Goriachuk:2018cac}, that the relation (\ref{eq:NSVZ_SQED}) stays valid in the class of the
renormalization prescriptions, which are called the NSVZ-type schemes, following the terminology of \cite{Jack:1996vg}.

When the theory is regularized with the help of higher derivatives (HD) \cite{Slavnov:1971aw,Slavnov:1972sq}
(see \cite{Faddeev:1980be} as well) in their supersymmetric form \cite{Krivoshchekov:1978xg,West:1985jx},
the equation (\ref{eq:NSVZ_SQED}) is valid in all orders of the perturbation theory (PT)
in terms of the bare coupling constant $a_0$ \cite{Stepanyantz:2011jy}.
This was explicitly verified in the 3-loop approximation in \cite{Stepanyantz:2011jy},
and also in \cite{Kazantsev:2014yna} within a bit different approach.
For this equation to be satisfied in the renormalized language, special boundary conditions $a=a_0$ and $Z = 1$,
imposed at a fixed value of $\mu$, were formulated in \cite{Kataev:2013eta}.
In the case of $\mu=\Lambda$ this scheme is called the procedure of minimal subtraction of logarithms (MSL).
In \cite{Nartsev:2016mvn} it was demonstrated that the HD + MSL prescription also provides
the NSVZ-type relation in renormalized softly broken SQED.
Further on it was shown in \cite{Kataev:2019olb} that the relation (\ref{eq:NSVZ_SQED}) is also satisfied
exactly in all orders of the PT in the on mass shell (OS) subtraction scheme
(used previously in SQED in \cite{Smilga:2004zr}).

More frequently for the regularization of SUSY models, instead of the dimensional regularization \cite{tHooft:1972tcz},
the method of dimensional reduction (DRED) is used \cite{Siegel:1979wq}.
The renormalization procedure in this case is the $\DRbar$ prescription, which is analogous to the $\MSbar$ scheme.
Its application leads to the violation of the expression (\ref{eq:NSVZ_SQED}) for the exact $\beta$-function
starting from the three-loop approximation \cite{Jack:1996vg}.
As was mentioned above, equation (\ref{eq:NSVZ_SQED}) can be restored using the finite renormalization
of the coupling constant, which should be fine tuned at every subsequent order of the PT.
In general, this redefinition may be fixed by the boundary conditions, analogous to the MSL-type procedure,
applied at the 3-loop order in \cite{Aleshin:2016rrr}.
The possibility of restoring the NSVZ relation in SQED, regularized by DRED,
was discussed earlier in \cite{Aleshin:2015qqc}.

In the case of the renormalized $\mathcal{N}=1$ SUSY Yang-Mills theory without matter the exact $\beta$-function
has the geometric series related form \cite{Jones:1983ip}
\begin{equation}
\label{eq:NSVZ-SYM}
\beta\left(a_s\right) = \frac{- 3 C_2 {a_s}^2}{4 - 2 C_2 a_s},
\end{equation}
where $C_2$ is the Casimir operator in the adjoint representation.

Let us clarify, how Eq. (\ref{eq:NSVZ-SYM}) was obtained.
In SUSY theories the related to axial and conformal anomalies operators should enter the same
supermultiplet and ought to be renormalized in the same way.
It is known, that the renormalization of the trace of the energy-momentum tensor is proportional
to the conformal anomaly factor $\left(\beta(a_s)/a_s\right)$, while the axial anomaly operator
should stay non-renormalized due to the Adler-Bardeen theorem \cite{Adler:1969er},
which is formulated for SUSY gauge models as well \cite{Jones:1982zf}.
On the first glance these facts are difficult to reconcile (see, e.g., \cite{Vainshtein:1985ed} and references there).
However, in \cite{Jones:1983ip} it was shown that they are consistent with each other when the the $\beta$-function
is given by (\ref{eq:NSVZ-SYM}).

The expression (\ref{eq:NSVZ-SYM}) does not agree with the 3-loop result \cite{Avdeev:1981ew,Velizhanin:2008rw},
obtained in the $\DRbar$ scheme.
They start to agree with each other after fixed in \cite{Jack:1996vg} at the 3-loop level finite renormalization
of the coupling constant, which restores equation (\ref{eq:NSVZ-SYM}).
There are strong indications that the identical expression for the $\beta$-function is valid
when the HD + MSL renormalization prescription is used \cite{Stepanyantz:2016gtk}.
As it was shown in \cite{Jones:1983ip}, the equation (\ref{eq:NSVZ-SYM}) resolves
the considered in \cite{Vainshtein:1985ed} anomaly puzzle of SUSY theories.
Another solution to this puzzle was proposed in \cite{Kazakov:1984bh}.

In this paper it is demonstrated that there is a class of renormalization schemes
in $\mathcal{N}=1$ SUSY Yang-Mills theory without matter, in which the expression (\ref{eq:NSVZ-SYM}) is exactly valid.
The group structure of the transformations acting in this class is investigated.
Their analogy to the finite renormalizations, conserving the $\beta$-function of non-supersymmetric
quantum chromodynamics (QCD) in the C-scheme, is considered.

\vspace{12pt}\textbf{2.}
Let us remind, that in $\mathcal{N}=1$ SQED the change of the renormalization scheme is performed in the following way:
\begin{equation}
\label{eqs:SQEDfin-ren}
a^\prime\left(a_0,~\mu/\Lambda\right) = a^\prime\left(a\left(a_0,~\mu/\Lambda\right)\right), \quad
Z^\prime\left(a^\prime\left(a\right),~\mu/\Lambda\right) = z\left(a\right)Z\left(a,~\mu/\Lambda\right),
\end{equation}
where $Z$ and $Z^\prime$ are the renormalization constants of matter in the schemes under consideration,
while the finite functions $a^\prime\left(a\right)$ and $z\left(a\right)$ can be chosen arbitrarily.
Choosing the renormalization procedures of the equations (\ref{eqs:SQEDfin-ren}) whithin the class of NSVZ schemes,
one arrives to the following condition \cite{Goriachuk:2018cac}:
\begin{equation}
\frac{1}{a^\prime\left(a\right)} - \frac{1}{a} - N_f \ln{z\left(a\right)} = \pi B =
- \frac{N_f}{2} \ln{\frac{{\mu^\prime}^2}{\mu^2}}, \label{eq:NSVZ_SQEDcons_cond}
\end{equation}
which is valid in all orders of the PT and the parameter $B$ doesn't depend on $a$.
This condition relates the particular representatives of the class of NSVZ schemes: HD + MSL \cite{Kataev:2013eta},
HD + OS \cite{Kataev:2019olb}, DRED + $\DRbar$ + (special finite renormalization) \cite{Aleshin:2016rrr} prescriptions
among others.

In general, the transformations within the class of all these schemes are parameterized by the set
$\Big\lbrace a^\prime\left(a\right)$, $z\left(a\right)$, $B\Big\rbrace$, where
$a^\prime\left(a\right)$ and $z\left(a\right)$ conserve the form of the exact $\beta$-function (\ref{eq:NSVZ_SQED}).
The function $z\left(a\right)$, performing the finite renormalization of matter, and the variable $B$
are convenient to choose as the independent quantities.

Consider now two transformations, parameterized by the sets
$\Big\lbrace a_i\left(a\right)$, $z_i\left(a\right)$, $B^{(i)}\Big\rbrace$ (with $i=1,2$),
which satisfy the restriction (\ref{eq:NSVZ_SQEDcons_cond}).
Note, that the second order expansion of the function $z_i\left(a\right)$ has the form
\begin{equation}
z_{i}\left(a\right) = 1 + D_1^{(i)} a + D_2^{(i)} a^2 + O\left(a^3\right), \label{eqs:NSVZ_SQEDcons_z}
\end{equation}
while the related PT expansion for $a_i\left(a\right)$ can be found from the following equation:
\begin{equation}
\frac{1}{a_{i}\left(a\right)} = \frac{1}{a} + \pi B^{(i)} + N_f \ln{z_{i}\left(a\right)}. \label{eqs:NSVZ_SQEDcons_a}
\end{equation}
Thus, to describe the transformations within the considered class in the three-loop approximation
it is necessary to fix three coefficients, namely $B^{(i)}$, $D_1^{(i)}$ and $D_2^{(i)}$.

It was shown in \cite{Goriachuk:2018cac}, that transformations, discussed above,
constitute a subgroup of general renormalization group transformations.
Indeed, parameterized by the sets $\Big\lbrace a_1\left(a\right)$, $z_1\left(a\right)$, $B^{(1)}\Big\rbrace$ and
$\Big\lbrace a_2\left(a\right)$, $z_2\left(a\right)$, $B^{(2)}\Big\rbrace$ finite renormalizations form
a composition, which satisfies the following condition
\begin{equation}
\label{eq:composition1}
a^\prime\left(a\right) = a_2\left(a_1\left(a\right)\right), \quad
z\left(a\right) = z_2\left(a_1\left(a\right)\right) z_1\left(a\right), \quad
B = B^{(2)} + B^{(1)},
\end{equation}
and provides the validity of (\ref{eq:NSVZ_SQEDcons_cond}) as well.
For every transformation, characterized by the set of functions
$\Big\lbrace a^\prime\left(a\right)$, $z\left(a\right)$, $B\Big\rbrace$, there is an inverse one, namely:
\begin{equation}
\label{eqs:RG_inverse}
\left\lbrace{a\left(a^\prime\right), \quad 1/z\left(a\left(a^\prime\right)\right), \quad - B}\right\rbrace.
\end{equation}
By definition, the identical element in the group of finite renormalizations is
\begin{equation}
\label{eqs:RG_unit}
a^\prime\left(a\right) = a, \quad z\left(a\right) = 1, \quad B = 0.
\end{equation}
One can verify that the finite renormalizations (\ref{eqs:RG_inverse}-\ref{eqs:RG_unit})
satisfy the restriction (\ref{eq:NSVZ_SQEDcons_cond}).
Similarly to the equations (\ref{eqs:NSVZ_SQEDcons_z}) and (\ref{eqs:NSVZ_SQEDcons_a}) one can write down the PT
expansion for the functions $z\left(a\right)$, $a^\prime\left(a\right)$ and
find the inverse one for the latter, i.e. $a\left(a^\prime\right)$.

Note, that in the case of $\mathcal{N}=1$ SQED the given subgroup is non-commutative.
This implies the result of the following composition of transformations
\begin{equation}
\label{eq:composition2}
a^\prime\left(a\right) = a_1\left(a_2\left(a\right)\right), \quad
z\left(a\right) = z_1\left(a_2\left(a\right)\right) z_2\left(a\right), \quad
B = B^{(1)} + B^{(2)}
\end{equation}
doesn't coincide with the result of the finite renormalization (\ref{eq:composition1}).
To verify this consider the compositions (\ref{eq:composition1}) and (\ref{eq:composition2}) in the three-loop
approximation and substitute the expansion (\ref{eqs:NSVZ_SQEDcons_z}) in both of them,
employing the formula (\ref{eqs:NSVZ_SQEDcons_a}).
The coincidence of these compositions in the 3-loop order is only achieved when the coefficients
in (\ref{eqs:NSVZ_SQEDcons_z}) and (\ref{eqs:NSVZ_SQEDcons_a}) are related as
\begin{equation}
B^{(1)} D_{1}^{(2)} = B^{(2)} D_{1}^{(1)}.
\end{equation}
In general the coefficients $B^{(i)}$ and $D_1^{(i)}$ are arbitrary and should not satisfy this condition.
Moreover, in the next orders of the PT the similar conditions appear, including higher order coefficients
of $z_i\left(a\right)$ and $a_i\left(a\right)$ expansions (e.g., $D_2^{(i)}$ in equation (\ref{eqs:NSVZ_SQEDcons_z})).
For this reason the compositions (\ref{eq:composition1}) and (\ref{eq:composition2}) provide different results
and thus the subgroup, conserving equation (\ref{eq:NSVZ_SQED}) in $\mathcal{N}=1$ SQED,
in general, is non-commutative (non-Abelian).

\vspace{12pt}\textbf{3.}
Consider now $\mathcal{N}=1$ SUSY Yang-Mills theory without matter.
In this theory there are also transformations, which conserve the form of its exact $\beta$-function,
given in (\ref{eq:NSVZ-SYM}).
They are defined by finite renormalizations of the coupling constant ${a_s}^\prime\left({a_s}\right)$,
which in analogy to $\mathcal{N}=1$ SQED equation (\ref{eq:NSVZ_SQEDcons_cond}) satisfy the following condition
\begin{equation}
\frac{1}{{a_s}^\prime\left(a_s\right)} - \frac{1}{a_s} +
\frac{\beta_{1}}{\beta_{0}} \ln{z_{\alpha}\left(a_s\right)} =
\pi \tilde{B} \equiv \beta_{0} \ln{\frac{{\mu^\prime}^2}{\mu^2}}, \label{eq:NSVZ_SYMcons_cond}
\end{equation}
where $z_{\alpha}\left(a_s\right) \equiv {a_s}^\prime\left(a_s\right)/a_s$ and the first two coefficients
of the $\beta$-function (\ref{eq:NSVZ-SYM}), calculated in \cite{Jones:1980fx}, read:
\begin{equation}
\beta_0 = \frac{3}{4} C_2, \qquad \beta_1 = \frac{3}{8} {C_2}^2. \label{eq:beta0beta1_SYM}
\end{equation}
Comparing the restrictions (\ref{eq:NSVZ_SQEDcons_cond}) and (\ref{eq:NSVZ_SYMcons_cond})
we conclude that the latter is more strict.
Indeed, it contains only one unfixed parameter $\tilde{B}$,
which is similar to $B$ in equation (\ref{eq:NSVZ_SQEDcons_cond}).
It is enough to chose this parameter in an arbitrary way to define the finite renormalization
${a_s}^\prime\left(a_s\right)$ unambiguously.

Finite renormalizations, defined by the condition (\ref{eq:NSVZ_SYMcons_cond}),
correspond to the change of the scale $\mu$ and form a one-parameter commutative (Abelian) subgroup
of general renormalization group transformations. Let us verify this explicitly.
The composition of sequential renormalizations $\Big\lbrace {a_s}_1\left(a_s\right)$, $\tilde{B}^{(1)}\Big\rbrace$ and
$\Big\lbrace {a_s}_2\left(a_s\right)$, $\tilde{B}^{(2)}\Big\rbrace$, which satisfy the condition
(\ref{eq:NSVZ_SYMcons_cond}), does not depend on the order of action:
\begin{equation}
{a_s}^\prime\left(a_s\right) = {a_s}_2\left({a_s}_1\left(a_s\right)\right) = {a_s}_1\left({a_s}_2\left(a_s\right)\right),
\end{equation}
and provides the validity of (\ref{eq:NSVZ_SYMcons_cond}) with $\tilde{B} = \tilde{B}^{(1)} + \tilde{B}^{(2)}$.
The neutral element of the identical transformation
$\Big\lbrace{a_s}^\prime\left(a_s\right) = a_s$, $\tilde{B} = 0\Big\rbrace$ belongs to the studied subgroup.
For an arbitrary finite renormalization, there is an inverse transformation
$\Big\lbrace a_s\left({a_s}^\prime\right)$, $-\tilde{B}\Big\rbrace$, which is analogous to (\ref{eqs:RG_inverse}).
Thus, the set of finite renormalizations, satisfying the restriction (\ref{eq:NSVZ_SYMcons_cond}), is
a commutative subgroup in contrast to the non-commutative one, discussed in section 2.

Note also that the corresponding to the change of $\mu$ transformation properties of reflexivity,
transitivity and symmetry were considered earlier in \cite{Brodsky:2012ms}.
In the case of $\mathcal{N}=1$ SUSY Yang-Mills theory the subgroup of these transformations
is theoretically distinguished, since they conserve the form of the exact $\beta$-function (\ref{eq:NSVZ-SYM}).

\vspace{12pt}\textbf{4.}
In QCD the special C-scheme has recently been proposed \cite{Boito:2016pwf}.
It was used for example in \cite{Jamin:2017mul,Baikov:2018wgs,Davies:2017hyl}
to study the analytical contributions to QCD perturbative series of the terms, proportional to Riemann functions
$\zeta\left(n\right)$ with even integer arguments ($n = 4, 6, \dots$).
By definition, the $\beta$-function in the C-scheme \cite{Boito:2016pwf} is
\begin{equation}
\beta\left(a_s\right) = \frac{-\beta_0 {a_s}^2}{1 - \left(\beta_1/\beta_0\right) a_s}. \label{eq:betaC}
\end{equation}
The coefficients $\beta_0$ and $\beta_1$ were evaluated in \cite{Gross:1973id,Politzer:1973fx} and
\cite{Jones:1974mm,Egorian:1978zx} respectively. They read
\begin{equation}
\beta_{0} = \frac{11}{12} C_2 - \frac{1}{3} T_F n_f, \quad
\beta_{1} = \frac{17}{24} {C_2}^2 - \frac{5}{12} C_2 T_F n_f - \frac{1}{4} C_F T_F n_f, \label{eq:beta0beta1_QCD}
\end{equation}
where $n_f$ is the number of quark flavours, $C_F$ and $C_2$ are the Casimir operators in the fundamental and adjoint
representations of the gauge group, $T_F$ is the Dynkin index.

The expression (\ref{eq:betaC}) is consistent with the equation (\ref{eq:NSVZ-SYM}) if instead of the coefficients
$\beta_0$ and $\beta_1$ given in (\ref{eq:beta0beta1_QCD}) one takes their values (\ref{eq:beta0beta1_SYM})
for $\mathcal{N}=1$ SUSY Yang-Mills theory.
In the latter case $\beta_{1}$ is divided by $\beta_{0}$ in the denominator of (\ref{eq:betaC}) without the remainder.
Moreover, the same feature also takes place in QCD without quarks (gluodynamics),
when $\beta_0$ and $\beta_1$ are defined by the analog of (\ref{eq:beta0beta1_QCD}) with $T_F n_f = 0$.

To relate the renormalization group quantities in the $\MSbar$ and the C-scheme,
it is required to perform the following finite renormalization:
\begin{equation}
{a_s}^\prime = a_s + \left(\frac{{\beta_1}^2}{{\beta_0}^2} - \frac{\beta_2}{\beta_0}\right) {a_s}^3 +
O\left({a_s}^4\right). \label{eq:fromAny-toC}
\end{equation}
In the case of the $\MSbar$ scheme $\beta_{2}$ was analytically computed in \cite{Tarasov:1980au,Larin:1993tp}.
In gluodynamics the corresponding expression is $\beta_{2} = \left(2857/3456\right) {C_2}^3$
and is divided by $\beta_0$ in (\ref{eq:fromAny-toC}) without the remainder,
similarly to the case of $\mathcal{N} = 1$ SUSY Yang-Mills theory without matter.
Therefore, the transformation (\ref{eq:fromAny-toC}) is analogous to the renormalization ${a_s}^\prime\left(a_s\right)$,
which leads to the exact $\beta$-function (\ref{eq:NSVZ-SYM}) from the $\DRbar$ scheme result,
and the C-scheme $\beta$-function in gluodynamics is analogous to the exact $\beta$-function (\ref{eq:NSVZ-SYM}) itself.

Now return to to the case of QCD with quarks.
Any finite renormalizations, which conserve the form of (\ref{eq:betaC}), satisfy the exact restriction:
\begin{equation}
\frac{1}{a_s^\prime} - \frac{1}{a_s} + \frac{\beta_1}{\beta_0} \ln{\frac{a_s^\prime}{a_s}} = \beta_{0} C \equiv
\beta_{0} \ln{\frac{{\mu^\prime}^2}{\mu^2}}. \label{eq:Ccons_cond}
\end{equation}
The similar expression was derived in \cite{Boito:2016pwf},
where the appearance of the parameter $C$ explains the term ``C-scheme".
In $\mathcal{N}=1$ SUSY Yang-Mills theory the formula (\ref{eq:NSVZ_SYMcons_cond}) has the same form and meaning as
the condition (\ref{eq:Ccons_cond}), and the parameters $\tilde{B}$ and $C$ are related:
\begin{equation}
\tilde{B} = \frac{3 C_2}{4\pi} C.
\end{equation}

Note that the variable $\Delta$, analogous to $C$, was introduced earlier in \cite{Vladimirov:1979my}
while solving the renormalization group equations for the two-loop $\beta$-function
of a general asymptotically-free theory.
In the case of QCD these variables are identical:
\begin{equation}
\Delta = \ln{\frac{{\mu^\prime}^2}{\mu^2}} = C.
\end{equation}
The difference is that in the C-scheme the $\beta$-function is defined by formula (\ref{eq:betaC}) and
the corresponding renormalization group equations are solved in higher orders of the PT.

Finite QCD renormalizations, which satisfy the condition (\ref{eq:Ccons_cond}),
are analogous to those considered in section 3 in the case of $\mathcal{N}=1$ SUSY Yang-Mills theory.
They form the commutative subgroup of general renormalization group transformations too.
This subgroup is entirely characterized by the variable C.

\section*{Conclusion}
In this paper it has been demonstrated, that in $\mathcal{N}=1$ SUSY Yang-Mills theory without matter superfields
there is a class of renormalization schemes,
in which the exact $\beta$-function (\ref{eq:NSVZ-SYM}) is valid in terms of the renormalized coupling constant.
Acting within this class transformations correspond to the change of the scale $\mu$ and
form a one-parameter commutative subgroup of finite renormalizations.
It has been shown, that the analogous transformations in QCD conserve the form of the C-scheme $\beta$-function.
It has also been found, that conserving the equation (\ref{eq:NSVZ_SQED}) subgroup of finite renormalizations
in $\mathcal{N}=1$ SQED, in general, is non-commutative.

\section*{Acknowledgments}
This work continues the research done in \cite{Goriachuk:2018cac,Kataev:2019olb,Aleshin:2016rrr}.
The authors are grateful to K.~V.~Stepanyantz for the interest to this work and useful discussions.
This work was supported by the Foundation for Advancement of Theoretical Physics and Mathematics ``Basis",
grant No.~17-11-120.

\begin{flushleft}

\end{flushleft}
\end{document}